\documentclass[%
 aip,
 amsmath,amssymb,
 reprint,%
]{revtex4-1}

\usepackage{graphicx}
\usepackage{dcolumn}
\usepackage{bm}

\usepackage[utf8]{inputenc}
\usepackage[T1]{fontenc}
\usepackage{mathptmx}
\usepackage{etoolbox}

\makeatletter
\def\@email#1#2{%
 \endgroup
 \patchcmd{\titleblock@produce}
  {\frontmatter@RRAPformat}
  {\frontmatter@RRAPformat{\produce@RRAP{*#1\href{mailto:#2}{#2}}}\frontmatter@RRAPformat}
  {}{}
}%
\makeatother

\newcommand{\beginsupplement}{
    \onecolumngrid
    \setcounter{table}{0}
    \renewcommand{\thetable}{S\arabic{table}}
    \setcounter{figure}{0}
    \renewcommand{\thefigure}{S\arabic{figure}}
    \setcounter{equation}{0}
    \renewcommand{\theequation}{S\arabic{equation}}
}

\begin{document}

\preprint{AIP/123-QED}


\title
{Sn/InAs Josephson junctions on selective area grown nanowires with in-situ shadowed superconductor evaporation}
\author{A. Goswami}
\email{aranyagoswami9@gmail.com}
\affiliation{Electrical and Computer Engineering Department, University of California, Santa Barbara, Santa Barbara, CA 93106,USA
}%
\author{S.R.Mudi}%
\affiliation{ 
Department of Physics and Astronomy, University of Pittsburgh, Pittsburgh, PA 15260, USA
}%

\author{C. Dempsey}
\affiliation{%
Electrical and Computer Engineering Department, University of California, Santa Barbara, Santa Barbara, CA 93106,USA
}%

\author{P. Zhang}%
\affiliation{ 
Department of Physics and Astronomy, University of Pittsburgh, Pittsburgh, PA 15260, USA
}%

\author{H.Wu}%
\affiliation{ 
Department of Physics and Astronomy, University of Pittsburgh, Pittsburgh, PA 15260, USA
}%

\author{W.J.Mitchell}%
\affiliation{ 
Nanofabrication facility, University of California, Santa Barbara, Santa Barbara, CA 93106,USA
}%

\author{S.M.Frolov}%
\affiliation{ 
Department of Physics and Astronomy, University of Pittsburgh, Pittsburgh, PA 15260, USA
}%

\author{C.J. Palmstrom}
\email{cjpalm@ucsb.edu}
\affiliation{%
Electrical and Computer Engineering Department, University of California, Santa Barbara, Santa Barbara, CA 93106,USA
}%
\affiliation{%
Materials Department, University of California, Santa Barbara, Santa Barbara, CA 93106,USA
}%

\date{\today}


\maketitle

\begin{quotation}
Abstract : Superconductor-semiconductor nanowire hybrid structures are useful in fabricating devices for quantum information processing. While selective area growth (SAG) offers the flexibility to grow semiconductor nanowires in arbitrary geometries, in-situ evaporation of superconductors ensures pristine superconductor-semiconductor interfaces, resulting in strong induced superconductivity in the semiconducting nanowire. In this work, we evaporated islands of superconductor tin on InAs SAG nanowires, by using in-situ shadowing with high aspect-ratio pre-fabricated SiO$_x$ dielectric walls. Our technique allows complete customization of each physical parameter of such hybrid nanostructures, while performing the nanowire  and superconductor growths without breaking vacuum. Using this technique, we grew super(S)-normal(N)-super(S) (SNS), NS and SNSNS junctions. We performed cryogenic electron transport measurements revealing the presence of gate and field tunable supercurrents in shadow junctions fabricated on in-plane SAG nanowires. We further measured the superconducting gap and critical fields in the hybrid nanostructures and the crossover from 2e to 1e periodicity in the SNSNS junctions, as a proof of the usability of these hybrid nanostructures.


\end{quotation}

\section{\label{sec:level1}Introduction
}




Nanostructures containing superconductor-semiconductor (super-semi) junctions are particularly interesting systems to study various uncommon quantum mechanical effects in electron transport. Particularly in topological physics, such hybrid systems comprising of s-wave superconductors in contact with semiconducting nanowires, have gained considerable interest in the search for the existence of Majorana modes in solid state systems\cite{Mourik2012SignaturesDevices}. Simultaneously, these super-semi nanowires are important for quantum information processing for realizing gate tunable superconducting qubits\cite{Larsen2015Semiconductor-Nanowire-BasedQubit, DeLange2015RealizationElements}, Cooper pair splitters\cite{Baba2018Cooper-pairNanowires,Kurtossy2022ParallelRepulsion} and quantum interference devices\cite{Heedt2021Shadow-wallDevices}. To enable these technologies, it is crucial to be able to fabricate high quality super-semi hybrid nanostructures. 

\begin{figure*}
\includegraphics[width=6in]{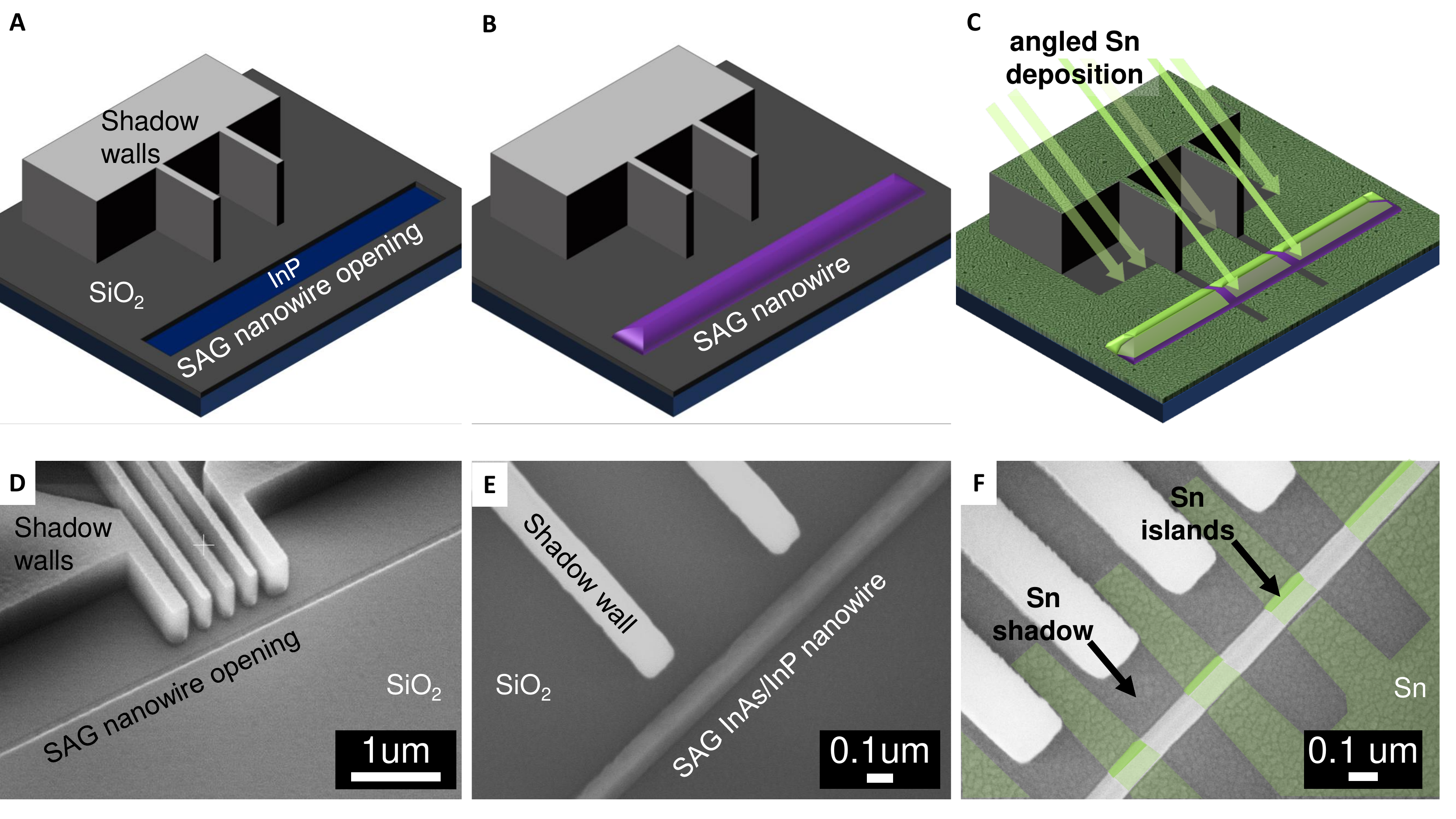} 
\centering
\caption{\label{fig:wide}The combined selective area growth and shadow deposition technique to fabricate super-semi hybrid structures. (A) shows the initial dielectric template with a shadow wall in close proximity to a SAG nanowire trench. The height of the shadow structures is  650$\mu$m. (B) demonstrates the selective area growth of the InAs nanowire as the first step of the process. There is no parasitic nucleation on any of the shadow walls. (C) shows the angled evaporation of the superconductor Sn on the nanowire, creating multiple islands of Sn with uncovered InAs in between. (D), (E) and (F) are scanning electron microscopy images corresponding to the stages in schematics (A), (B) and (C) respectively. Green regions in (F) are parts of the sample that are covered with Sn. }
\end{figure*}

To fabricate functional hybrid super-semi nanostructures, first, the semiconductor nanowire must contain low defect-densities to enable phase-coherent transport from one superconducting lead to the other. Growing complex networks of such nanowires can be achieved through selective area growth.
Simultaneously, it is also crucial that the interface between the superconductor and semiconductor is pristine and defect-free. This is essential in creating a transparent interface which enables Andreev reflection of electrons across the normal(semiconductor)-superconductor interface to proximitize the semiconducting channel. 

In past experiments, it has been demonstrated that if the growth of the nanowire and the deposition of superconductors are performed without breaking vacuum in between the steps, a sharp disorder free super-semi interface can be formed and the nanowire exhibits a clean proximitized superconducting gap without sub-gap states\cite{Chang2015HardNanowires,Carrad2020ShadowHybrids,Kanne2021EpitaxialDevices}. Deposition of superconductors shadowed by pre-patterned or grown nano-structures \cite{Heedt2021Shadow-wallDevices} offers flexibility in creating such superconducting junctions/islands in specific regions of the semiconducting nanowire, enabling the fabrication of complex multi-junction devices. Additionally, this technique avoids any post-growth etching of the superconductors on top of the semiconductor, which can prevent semiconductor damage from dry/wet etching processes, as well as, allow the use of a wider range of superconductors for which selective etch chemistries are unknown. 

There exists only 2 reports of using shadow wall superconductor deposition on in-plane selective area grown nanowires\cite{Jung2021UniversalNetworksb,Jiang2022SelectiveSubstrateb}. However, no extensive electrical transport characterizations were performed in these studies and most importantly, no evidence of tunable supercurrents was reported in these nanowires. In this work, we have demonstrated an integrated and highly customizable platform where superconducting tin (Sn) was deposited on selective area grown (SAG) in-plane InAs nanowires grown using chemical beam epitaxy (CBE). The Sn was deposited using angled evaporation and shadowing with pre-patterned dielectric wall structures. In particular, compared to previous works, the template fabrication method adopted in this work offers a high degree of flexibility in the height and aspect ratios of the shadow walls, and their proximities to the nanowire position. Using this technique, we demonstrated the fabrication of super(S)-normal(N)-super(S) (SNS), NS, SNSNS as well as longer chains of SNS junctions. We present transport experiments on the SNS junctions, revealing supercurrents of the order of hundreds of nanoAmperes that are tunable  by gate voltages and magnetic fields. We extracted proximitized superconducting gaps of approximately $560 \mu V$ in the InAs nanowire that were sizeable fractions of the bulk superconductor gap, as well as demonstrated 2e to 1e parity transitions in the SNSNS superconducting junctions. This demonstrates the feasibility of this one-shot growth and shadow-evaporation technique to fabricate Josephson junctions on in-plane SAG nanowires. In comparison to VLS nanowires, such in-plane hybrid SAG nanowires open up the possibility to explore more complex superconductor-semiconductor networks. Our work also presents the first reports of such transport measurements in the Sn-InAs nanowire hybrid system (to the best of our knowledge).

\section{\label{sec:level1}Experimental Details}

\subsection{\label{sec:level2}Fabrication of shadow wall tempaltes}

To fabricate the templates, first  a InP(001) substrate was masked with atomic layer deposited (ALD) aluminum oxide AlO$_x$ layer and plasma enhanced chemical vapor deposited (PECVD) SiO$_2$ layer. The AlO$_x$ layer protects the surface of the InP from plasma damage as well as non-volatile fluorine residues from dry etches. Following this, alignment marks were patterned using electron beam lithography (EBL) and reactive ion etching into the InP substrate using the SiO$_2$ and AlO$_x$ hard mask. All the deposited dielectrics were etched away and subsequently ALD AlO$_x$ and PECVD SiO$_2$ were redeposited. The SAG nanowire trenches were now exposed in the EBL and the SiO$_2$ layer was etched using inductively coupled plasma. An additional AlO$_x$ layer was deposited using ALD on the sample. This was followed by a 650nm thick layer of PECVD SiO$_2$ and a 150nm layer of ruthenium (Ru) deposited using sputtering.Hydrogen silsesquioxane (HSQ) was spun as a resist on top and the shadow wall patterns were exposed in the EBL and developed. The Ru layer was etched in an oxygen plasma using the HSQ mask. Then, the Ru (+exposed HSQ on top) mask was used to etch the SiO$_2$ shadow walls. This etch was performed using a CHF$_3$/CF$_4$ chemistry\cite{Mitchell2021HighlyMask}. This etch produces near vertical sidewalls and results in high aspect ratio shadow walls with very narrow fingers. Once the shadow walls were fully etched, the remaining Ru mask was etched off using an oxygen plasma and the AlO$_x$ layer was wet etched in AZ300MIF. A significant overtech (30-40\%) is required to ensure that the AlO$_x$ is removed from the narrow SAG trenches.Before loading into the CBE chamber each such shadow wall SAG sample underwent three cycles of digital etching (UV Ozone oxidation of exposed InP inside the trenches followed by wet etch in dilute HCl(1:20)). The digital etching was found to improve yield of nanowire growths inside the nanowire trenches. More details of the fabrication can be found in the Supplementary information.

\subsection{\label{sec:level2}Growth of nanowires and shadow deposition of tin}

The nanowires grown for these experiments have the layer structure InAs/InP buffer /InP (001) substrate. The InP buffer growth separates the transport channel from the residues and damages at that substrate interface, therefore improving the tranpsort characteristics of the device. The growth of both the InP buffer and the InAs channel were performed at 520°C (calibrated using the appearance of the (2x4) surface reconstruction in the reflection high energy electron diffraction signal of the InP(001) surface from a coloaded InP sample). Before initiating the InP buffer layer growth, a 3 minute anneal was performed at 520°C under P$_2$ overpressure.For the InP buffer layer growth, trimethylindium (TMI)  and thermally cracked phosphine (PH$_3$) were used and growth time was 8 minutes. A 3 minute anneal under cracked PH$_3$ was performed after the growth of the buffer layer to smoothen the InP surface. Thereafter, a 10 seconds anneal under thermally cracked AsH$_3$ was performed while switching the group-V precursors before turning on the TMI for the InAs growth. For InAs growth TMI  and cracked AsH$_3$ were used and growth time was 3 minutes. Following the growth of the nanowires, the samples were cooled down under overpressure of the cracked AsH$_3$. They were subsequently transferred into the superconductor deposition chamber with a liquid nitrogen cooled substrate holder with a base temperature of ~80K. The sample was kept in contact with the cold stage for approximately 1 hour for the temperature of the sample to equilibriate. Tin was then deposited on the sample at an angle of 40 degrees to the sample normal at a growth rate of 2 \AA/s from an effusion cell (maintained at a thermocouple temperature of 1100 °C). After the deposition of the tin, the sample was quickly (<10 minutes) transferred in-situ to the dielectric deposition chamber and 3nm of electron beam evaporated AlO$_x$ was deposited with sample rotation. This AlO$_x$ layer minimizes further oxidation of the Sn, when removed from the ultra-high vacuum environment.

Fig 1 shows the schematics of the process (A-C) and the corresponding scanning electron microscopy images (D-F) of a representative sample in each of the respective steps. The distance between the SAG nanowire trench and the shadow walls is less than 100nm (Fig 1E). This design thus presents the flexibility of the shadow deposition process where a longer superconductor shadow can be created using smaller angles between the evaporation cell and the sample normal. The InAs nanowire with InP buffer exhibit a triangular smooth top surface with high yield and negligible parasitic growth. This is particularly crucial, since it demonstrates that the AlO$_x$ layer provided enough protection from harsh etching conditions, to the starting surface of the SAG growth, resulting in a high yield of devices with this technique.  Post evaporation of the Sn, the Sn islands were formed on only one face of the SAG nanowire (Fig 1F). The high aspect ratio of the shadow walls enabled the formation of junctions with lengths of the order of ~100nm.

\subsection{\label{sec:level2}Device fabrication for transport measurements}
To study the induced superconducting properties of Sn on InAs nanowire, we fabricated two types of devices. The first were superconductor-nanowire-superconductor (S-NW-S) junctions (Fig 2-3). The second were island devices (Fig 4) with the Sn coated InAs island contacted on the two sides either by normal metal leads or superconducting leads and separated from the island by gate controlled tunnel barriers (N-NW-S-NW-N or S-NW-S-NW-S). Since Sn covers the entire substrate surface, in order to avoid shorting of the source-drain leads, first Sn was selectively etched from the SiO$_2$ surface around the nanowire. We first performed several etch tests to optimize the HCl solution concentration and the etch time. However, we found that Sn etched non-uniformly with either residues or unetched chunks of Sn left behind on the wafer. Additionally, there was always some overetching observed. The etch conditions can also vary from wafer to wafer or between different substrates \cite{zhang2022planar}. For the devices shown in this study, we found that a dilute 1:400 HCl:H$_2$O solution worked best for etching Sn. Prior to etching Sn, we defined windows around the shadow wall and nanowire using electron beam lithography with 950 PMMA A4 as the resist. For making devices, we selected nanowires with very little or no residues remaining around the junction after Sn etching. We then defined the source and drain leads for these nanowires using e-beam lithography and performed Ar ion milling to remove the AlO$_x$ cap prior to evaporation of the Ti/Au leads. For making the top gates, approximately 10 nm thick dielectric layer of hafmium oxide (HfO$_x$) was deposited using ALD followed by liftoff of Ti/Au top gates. Further details of the fabrication steps can be found in Supplementary Information.

\section{\label{sec:level1}Transport Measurements}

\begin{figure*}
\includegraphics[width=6in]{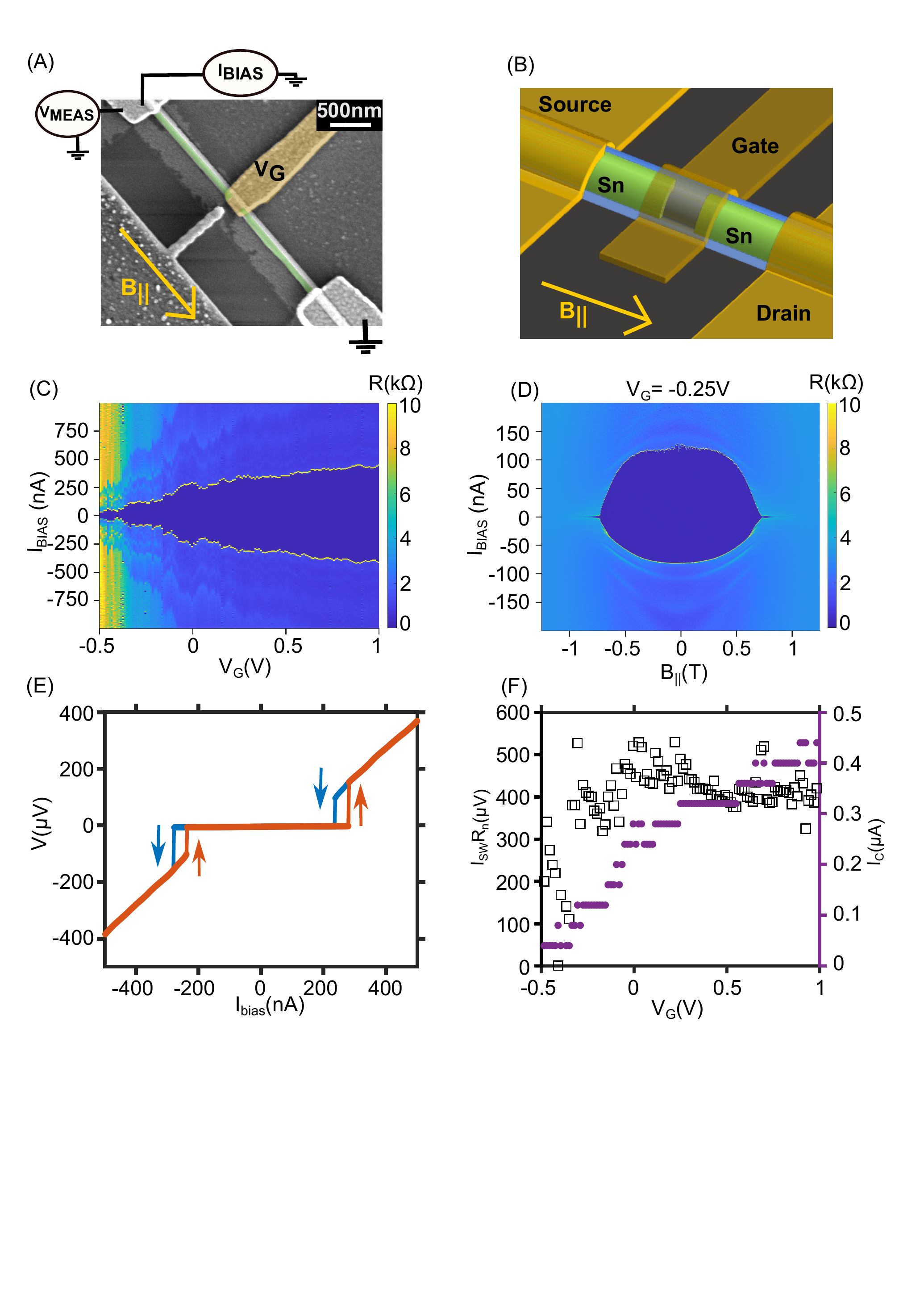} 
\centering
\caption{\label{fig:wide} Current bias measurements : (A) SEM image of the gated SNS device along with the measurement circuit. Sn shell on the nanowire is shown in green. Magnetic field $B_{||}$ is applied parallel to the nanowire and current is sourced through the wire while measuring voltage across it. (B) Schematic of (A). (C) Evolution of the differential resistance as a function of the gate voltage and $I_{BIAS}$. (D) Supercurrent evolution as a function of the in-plane magnetic field applied parallel to the nanowire. (E) I-V curve at $V_G$ = 0 V. The arrows represent the current sweep direction. (F) $I_{sw}R_{n}$ product (black squares) and $I_c$ (purple dots) as a function of top gate voltage. The step-like features in $I_{sw}$ arise due to the limited resolution from the high current bias scans.}
\end{figure*}

 \begin{figure*}
\includegraphics[width=6in]{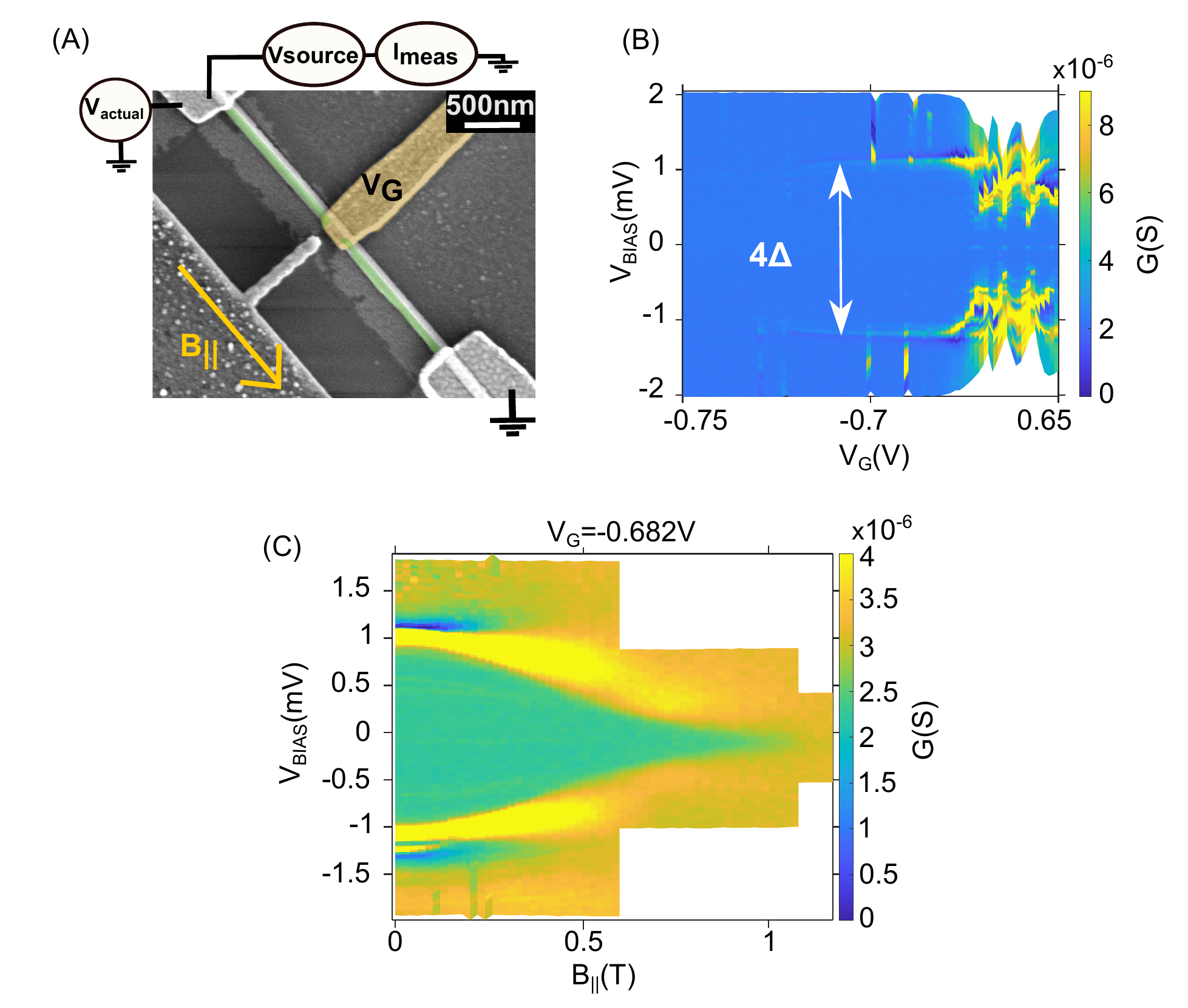} 
\centering
\caption{\label{fig:wide} Voltage bias measurements: (A) SEM image of the SNS junction from Fig. 2 with the measurement circuit for pseudo four-terminal $V_{bias}$ measurement. $V_{source}$ is the voltage applied using the voltage bias module connected in series with the current measurement module $I_{meas}$. $V_{actual}$ measures the fraction of the voltage applied across the nanowire in the presence of non-linear RC filters in our measurement setup. (B) Differential conductance as a function of $V_{BIAS}$ and top gate voltage $V_G$. The conductance peaks at finite bias are separated by $4\Delta$, which gives $\Delta = 560 \mu V$. (C) Differential conductance as a function $V_{BIAS}$ and magnetic field applied parallel to the nanowire.}
\end{figure*}

All transport measurements were performed in a dilution refrigerator at a base temperature of $\approx$ 40 mK. All the devices fabricated for this study have only two Ti/Au leads contacting the nanowires, hence we cannot perform 4 terminal measurements to eliminate line and contact resistances. Since our setup has non-linear RC filters, in order to remove their effect, we performed pseudo 4-terminal measurements by wire-bonding each Ti/Au lead to two different pads on the printed circuit board (This eliminates the line resistance but not the contact resistance). S-NW-S junctions were first measured in a DC current biasing configuration (Fig 2). From the differential resistance plot as a function of gate voltage and current bias (Fig 2(c)), we observed a gate tunable switching current $I_{sw}$. The magnitude of $I_{sw}$ decreased as the gate voltage was decreased and the electron density in the nanowire was depleted. 
We observed resonances 
above the switching current which likely arise due to multiple Andreev reflections (MARs), which are dips in resistance at constant voltage (See Supplementary Information Fig S3). The switching current exhibits a hysteresis in forward and reverse bias, possibly due to capacitive\cite{tinkham1996introduction} or local heating effects \cite{courtois2008origin,paajaste2015pb,gunel2012supercurrent,dartiailh2021missing}. This can be seen in the IV trace presented in Fig 2(e), at $V_g=0 V$. 

The $I_{sw}R_n$ product, calculated from the switching current $I_{sw}$ and the normal state resistance at high current bias $R_n$, is a figure of merit commonly used for characterizing Josephson junctions (Fig 2(f)). For $0 V < V_g <1 V$, this value ranged between 300-600 $\mu V$ for this device. This is a sizeable fraction of the superconducting gap of bulk tin (=600 $\mu V$ )and is comparable to previous reports in other similar systems \cite{pendharkar2021parity}. Some S-NW-S junctions among the devices measured demonstrated a much lower $I_{sw}R_n$ product (approximately 100 $\mu V$). This reduction in the $I_{sw}R_n$ product for some of the measured junctions could possibly be due to thermal activation leading to premature switching of the supercurrent or finite transparency of SC contacts \cite{tinkham1996introduction} or disorder \cite{Khan2020HighlyJunctions}.

Next, we studied the supercurrent evolution as a function of the magnetic field parallel to the nanowire (Fig 2(d)). For the device presented in Fig 2(a), we measured a critical parallel magnetic field of approximately 1 T for 15 nm thick Sn film on our nanowires. However, the critical fields of our other devices were in the range 1 - 1.8 T. This critical field value is similar to the value reported for S-NW-S junctions of the same thickness of Sn in InSb/Sn \cite{pendharkar2021parity}. We also observed different magnetic field dependence of the $I_{bias}$ vs axial magnetic field plots, specific to the nanowire measured and the gate voltage applied. In Fig 2(d), we see that for $B_{||}$ between $\approx \pm 0.5 T$, the junction switches from the superconducting to the resistive state at roughly the same $I_{sw}$. This could be resulting from premature switching of the supercurrent due to the resonances at finite bias. Additionally, for this device, we observe kinks at $B_{||} \approx$ 0.73 T and 0.8 T that do not change with gate voltage (see Fig S4). This cannot be explained in terms of interference effects, which was previously used to explain gate tunable nodes in magnetic field dependences of supercurrent \cite{zuo2017supercurrent}. 

\begin{figure*}
\includegraphics[width=6in]{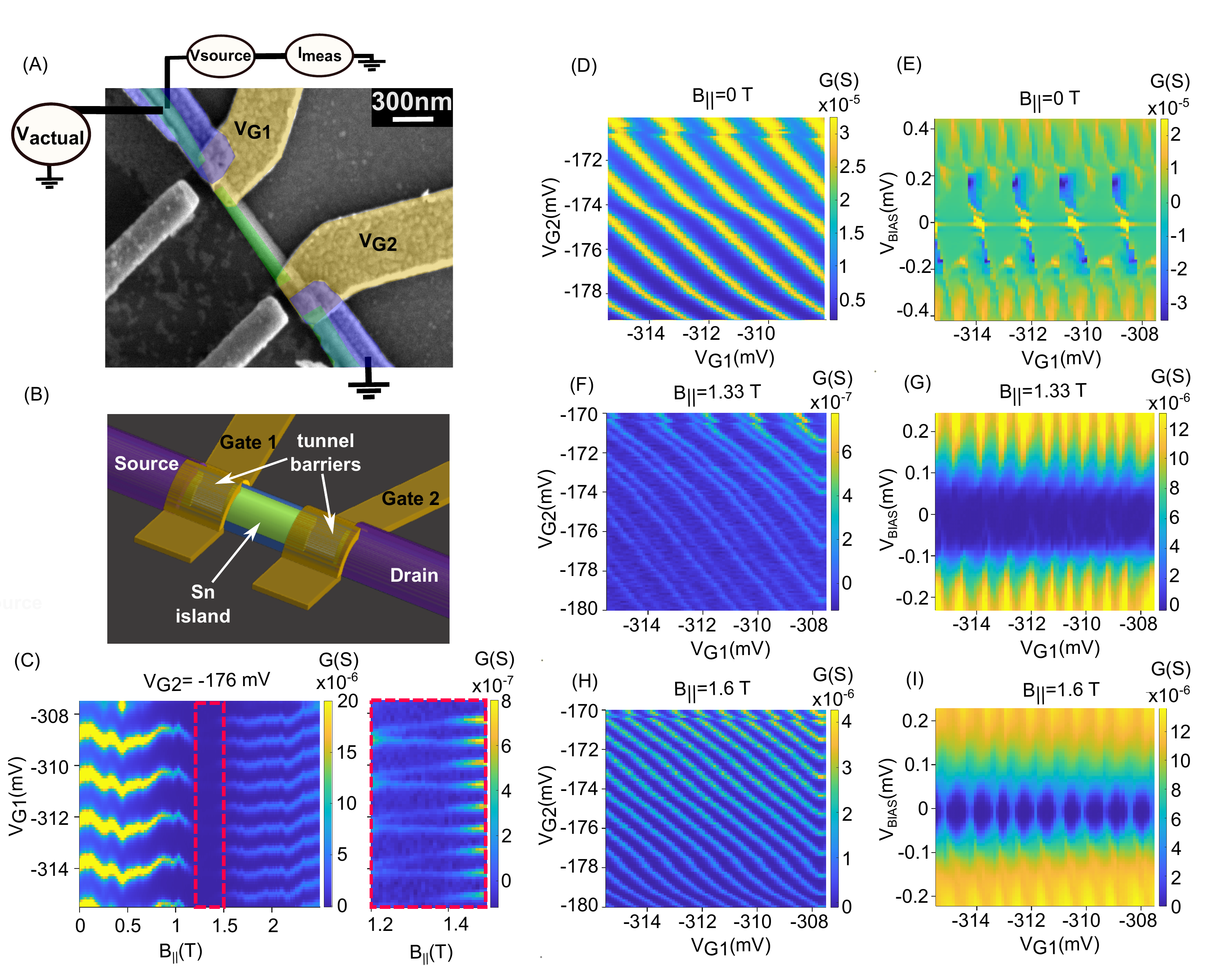} 
\centering
\caption{\label{fig:wide}Voltage bias measurements for the island device: (A) SEM image of an island device connected to normal metal leads (purple) via gate controlled tunnel barriers $V_{G1}$ and $V_{G2}$ along with the pseudo four terminal voltage bias circuit, similar to the $V_{bias}$ circuit used for the SNS junction in Fig 3. (B) schematic representation of (A). (C) Differential conductance as a function of $V_{G1}$ and axial magnetic field $B_{||}$ at $V_{G2}$ = -176 mV, $V_{DC}$ = 0 V and $V_{AC}$ = 10 $\mu$V. The 2e to 1e transition happens at $\approx 1.2 T$. Red dashed box shows a zoom-in of the transition region. (D, F, H) $V_{G1}$ vs $V_{G2}$ at $B_{||}$ = 0 T, 1.33 T and 1.6 T respectively. Plots were obtained at $V_{DC}$ = 0 V and $V_{AC}$ = 10 $\mu$V. (E, G, I) Differential conductance as a function of $V_{BIAS}$ and $V_{G1}$ at $B_{||}$ = 0 T, 1.33 T and 1.6 T respectively. $V_{G2}$ = -176 mV for all plots.}
\end{figure*}

The S-NW-S device was also measured in the voltage bias configuration. Fig 3(b) shows the differential conductance $dI/dV$ as a function of the actual source-drain voltage $V_{bias}$ and the applied top gate voltage $V_g$. For negative gate voltages ($V_g<-0.675 V$), the nanowire is in the pinched off regime and we see conductance peaks at finite bias due to tunneling between the superconductors. These two finite bias resonances are separated by 4$\Delta$ in $V_{bias}$, which gives an apparent superconducting gap of approximately 560 $\mu V$ - which is in good agreement with the bulk superconducting gap of Sn. This value of the apparent superconducting gap also agrees with the value for $\Delta$ obtained from tunnel spectroscopy of an NS device meausred in this study (see Fig S5(b)). For $-0.675 V < V_g < -0.65 V$, quantum dot features are observed, including Andreev bound states and supercurrent (faint zero bias peaks).

 Next, we studied the axial magnetic field dependence of this device under an applied voltage bias. Fig 3(c) shows the $V_{bias}$ vs $B_{||}$ plot with the gate voltage tuned to a value in the pinched off regime ($V_g = -0.682 V$).  
 The gap was observed to remain open till $\approx$ 1.1 T, which is higher than the critical field obtained from the $I_{bias}$ measurement in Fig 2(d). This value is less than the critical field of 3 T reported for 15 nm thin shells of Sn \cite{pendharkar2021parity}, although the gap does remain open upto 3 T in $V_{bias}$ for one of our devices (Fig S6 (f)). This variation in critical field values could be due to the nucleation of Sn on InAs being different compared to InSb, resulting in grains of different thicknesses on the nanowire (thicker grains can cause the superconductivity to be suppressed). Additionally, presence of a magnetic field component perpendicular to the Sn film on the nanowire can also cause a reduction in the value of the critical magnetic field. We estimate the misalignment of the magnetic field with the nanowire axis to be $\approx \pm 10 ^{\circ}$. For all the S-NW-S junctions measured in this study, we observed a higher critical field in $V_{bias}$ compared to the $I_{bias}$ measurements. We speculate that at very negative gate voltages, the nanowire is in the pinched off regime, and therefore we see the superconductivity evolution of only the Sn film on the nanowire, not of the induced superconductivity in the semiconducting segment. This is in contrast to the $I_{bias}$ measurements, which were obtained at less negative gate voltages, where the nanowire was still conducting.


The shadow wall deposition technique was also used to deposit an isolated tin island as demonstrated in Fig 4. The Sn island on the InAs nanowire was coupled to source drain leads via tunnel barriers, which were modulated using the two top gates $V_{g1}$ and $V_{g2}$ (Fig 4(a) and (b)). Figs 4(d), (f) and (h) show a gate versus gate scan at an applied lockin voltage of 10 $\mu V$ (dc voltage = 0 V) at three different magnetic field values parallel to the nanowire. The peak-dip structure seen in these plots arise from the Coulomb blockade effect, with the blockade being lifted at the peaks. The periodicity was halved at very high fields (1.6 T, Fig 4(h)) compared to the zero field plot (Fig 4(d)). At zero field, if the lowest quasiparticle energy state is above the charging energy of the island, then the island prefers to host only pairs of electrons in the ground state and transport through the island happens by tunneling of Cooper pairs. As the magnetic field is increased, the quasiparticle states move down in energy. At intermediate magnetic fields ($B_{||} = 1.33 T$), when the quasiparticle states are below the charging energy, the island alternates between even and odd parity states as the gate voltages are varied, which manifests as different peak spacings in Fig 4(f). For very high magnetic fields ($B_{||} = 1.6 T$, Fig 4(h)), the island is in the normal state and single electron transport sets in leading to the 1e periodicity.

The above three situations can also be visualised in the $V_{bias}$ vs $V_{g1}$ ($V_{g2}$ = -176 mV) plots shown in Fig 4(e), (g) and (f). At $B_{||} = 0 T$, for small $V_{bias}$, the island shows Coulomb diamonds which are 2e periodic in the gate voltage. Above $V_{bias} = 200 \mu V$, the periodicity in gate voltage is halved due to the onset of single electron transport through the island (Fig 4(e)). For $B_{||} = 1.33 T$ (Fig 4(g)), near zero bias, a second diamond of smaller width appears between the larger diamonds due to alternating even-odd parity. At $B_{||} = 1.6 T$, the island transitions to the normal state and the Coulomb diamonds become 1e periodic in gate voltage (Fig 4(g)) for all $V_{bias}$. The transition from 2e to 1e charging is also evident in the gate voltage versus magnetic field plot shown in Fig 4(c). We note that the critical field for the transition is approximately 1.2 T, which is higher than the values previously reported in other hybrid semiconductor-superconductor systems  \cite{albrecht2016exponential,Kanne2021EpitaxialDevices,shen2018parity,pendharkar2021parity}.


\section{\label{sec:level1}Conclusion}
In conclusion, through this work we first demonstrated a process to combine selective area growth of in-plane semiconductor nanowires and angled shadow deposition to create hybrid super-semi nanostructures. We utilized this technique to fabricate and measure Sn junctions of various geometries on in-plane SAG InAs nanowires with excellent yield. Next, through cryogenic transport measurements performed on S-NW-S junctions, we demonstrated the presence of a gate-controllable supercurrent and a superconducting gap in the InAs nanowire. Further, we performed measurements on Sn islands on InAs nanowires fabricated using the same technique, which exhibit the 2e to 1e parity transition with a high critical field. These measurements show that in-plane SAG nanowires with shadow evaporated superconductor is a viable platform to probe the physics of electron transport in hybrid nanostructures.

The technique allows the fabricaton of arbitrarily high aspect ratio shadow walls, defined solely by the thickness of the deposited silicon oxide. As demonstrated in this work, these shadow walls can be situated in close proximity to the SAG nanowire trenches enabling the formation of a long superconductor shadow which can aid in shadowing larger nanowire networks. Several such shadow walls can also be used to shadow multiple materials and create interesting interfaces in a fully in-situ and epitaxial method.In the future, this work therefore presents opportunities for fabricating a wide range of more complex super-semi hybrid nanowire networks and studying their transport physics.

\begin{acknowledgments}
We acknowledge the ANR-NSF PIRE:HYBRID OISE-1743717, the University of California, Santa Barbara (UCSB) National Science Foundation (NSF) Quantum Foundry through Q-AMASE-i Program via Award No. DMR-1906325 for the device fabrication, epitaxial growths, superconductor growths and microscopic studies. We also acknowledge the use of shared facilities of the UCSB Materials Research Science and Engineering Centers (No. NSF DMR 1720256) and the Nanotech UCSB Nanofabrication Facility. 
\end{acknowledgments}

\section*{Data Availability Statement}

Full data is available at 10.5281/zenodo.7688657

\section*{References}
\nocite{*}
\bibliography{ref_SW,ref_2}

\clearpage
\title{Supplementary Information}
\maketitle
\beginsupplement

\section*{Duration and Volume of Study}
The study lasted for 10 months, between January 2022 - October 2022. During this time, we fabricated 2 full 2" wafers with shadow wall templates. 5 different nanowire-superconductor growths were performed with several samples in each growth. Out of these, we fabricated two base chips. A total of nine S-NW-S junctions, six island devices and three NW-S junction on the two chips were measured over four cooldowns. Approximately 2500 data sets were collected.

\section*{Details of Fabrication}

Fabrication of the SAG nanowire/shadow-wall mask device on InP(001) substrates utilized 3 separate fabrication steps.  In all cases, a thin aluminum oxide (AlO$_x$) film was first deposited via atomic layer deposition (ALD) underneath all thicker SiO$_2$ films deposited via plasma-enhanced chemical vapor deposition (PECVD) in order to protect the underlying
substrate from plasma damage and non-volatile fluorine residues that result from the fluorine-based SiO$_2$ dry etches.

1.Alignment marks, etched into the InP substrate, were defined by first coating the substrate with a thin ALD AlO$_x$ film (6nm) and a thicker PECVD SiO$_2$ hard mask film (30 nm).  The alignment mark patterns were exposed using electron beam lithography into CSAR-62 positive resist (100 nm) and following development in MIBK/IPA solvents, the pattern was transferred into the SiO$_2$ hardmask using a CHF$_3$/CF$_4$/O$_2$ plasma in an inductively coupled plasma (ICP) etcher.  This dry etch stopped in the thin AlO$_x$ film and this film was removed by wet etching in AZ300MIF base developer which exposed the
InP substrate within the mark patterns.  A final dry etch step in a reactive ion etch (RIE) tool using CH$_4$/H$_2$/Ar chemistry transferred the mark patterns into the InP substrate to approximately 1$\mu$m depth.  The remaining SiO$_x$/AlO$_x$ films were then removed using a buffered oxide etch (BOE) and AZ300MIF dip.  Note that etched alignment marks, rather than the standard metal marks, were necessary for compatibility with the later chemical beam epitaxy (CBE) growth steps.

2.Generation of the SAG nanowire trenches required redeposition of the thin ALD AlO$_x$ (6 nm) film and PECVD SiO$_2$ (30 nm) film on the substrate.  The trench patterns were patterned using electron beam lithography into positive
CSAR-62 resist (100nm) and the trenches were etched into the SiO$_2$ layer using the same chemistry as in the alignment mark step above once again stopping on the protective AlO$_x$ layer, a layer that was kept intact.

3.To fabricate the shadow-wall mask, an additional 20 nm of ALD AlO$_x$ was deposited on the patterned substrate followed by a much thicker PECVD SiO$_2$ film (650 nm) and a 150 nm mask layer of Ru deposited using DC sputtering.  Negative hydrogen silesquioxane (HSQ) resist was used to define
the shadow-wall pattern using electron beam lithography (which was precisely positioned with respect to the nanowire trench patterns defined in the previous step using the alignment marks from the first step).  After development in 25 percent tetramethyl ammonium hydroxide solution, the HSQ pattern was transferred into the Ru mask layer using O$_2$/Cl$_2$ chemistry in an ICP etcher.  Ru has been shown to exhibit high selectivity in carbo-fluoro based etch chemistry and can produce near-vertical etch profiles in thick SiO$_2$ films \cite{Mitchell2021HighlyMask}.  Hence, the Ru pattern was used to mask the CHF$_3$/CF$_4$ ICP etch through the 650 nm thick SiO$_2$ film, a process which resulted in high aspect ratio shadow-walls with very narrow fingers and near-vertical sidewalls, all important requirements for the successful fabrication of the final device. Once completed, the Ru mask layer was removed by a short O$_2$/Cl$_2$ ICP etch and the remaining AlO$_x$ protective stop layer was removed by wet etching in AZ300MIF.  Note that an extended soak was required to ensure that the AlO$_x$ was completely removed from the narrow SAG nanowire trenches.

Before loading into the CBE chamber, the patterned shadow wall SAG samples were subject to three digital oxidation/etching cycles (each cycles consists
of a UV ozone oxidation step of the exposed InP surface inside the trenches immediately followed by a wet etch in dilute 1:10 HCl:H$_2$O solutions to remove
the surface oxide). The digital etching was found to significantly improve nanowire growth yields within the narrow trenches.

For device fabrication, as mentioned in the main text, the first step is to remove Sn from around the SAG nanowire structure to avoid any shorting between the leads. For this, we defined etch windows using two separate rounds of e-beam lithography (EBL) - first, for defining small windows around the shadow walls, and then writing large patterns for etching Sn under the leads only for those nanowires where no large Sn residues were visible near the shadow junctions. Each of the above EBL steps was followed by resist development, AlO$_x$ and Sn etching steps. For removing the AlO$_x$ cap, we used a 1:20 CD26:DI water solution for 2 minutes, followed by two consecutive 10 seconds and 50 seconds rinses in DI water. Dilute 1:400 HCl:DI water (36.5\%-38\% HCl) solution under magnetic stirring was used for 15 seconds to etch Sn, followed by two rinses for 10 seconds and 50 seconds in DI water. A nitrogen jet was used for blow drying the chip after each set of etch and rinse steps. A third round of EBL was done to define the source-drain leads. After developing the resist, 10nm Ti/120 nm Au was deposited using an e-beam evaporation system. Ar ion milling was done in steps for a total of 75 seconds at 15 mA and 250 V to remove the AlO$_x$ cap from the nanowire before depositing the leads. After a standard liftoff process in acetone, we deposited approximately 10 nm of HfO$_x$ in an atomic layer deposition system at 120\textdegree C. This was followed by a fourth round of EBL for writing top gates. 10nm Ti/150 nm Au was deposited without Ar ion milling for the gates. Prior to each EBL step above, the chip was spin-coated with e-beam resist 950 PMMA A4 and vacuum baked overnight. Vacuum baking was used to avoid potential dewetting of the cold-deposited Sn at high temperatures. For developing the resist, we use 1:3 MIBK:IPA solution for 1 minute followed by rinsing in IPA for 1 minute, then blow drying the chip with nitrogen.

\section*{Uncertainty in extracting $I_{sw}R_n$ product}
We have explored two different methods for extracting the $I_{sw}R_n$ product for SNS junctions - the threshold method and the peak finder method. In the threshold method, $I_{sw}$ is defined as the first current bias value where the differential resistance is non-zero, signaling a transition to the resistive state. In the peak finder method, $I_{sw}$ is equal to the current bias value at  which the first peak in differential resistance occurs. The threshold method typically gives a switching current which is lower than that obtained from the peak finder method; this in turn leads to lower $I_{sw}R_n$ product using the threshold method. This is illustrated in Fig S2.
\par For devices where the switching current is small (less than 100 nA in our case), the high current bias plots (current bias up to a few $\mu A$) used for extracting $I_{sw}$ and $R_n$ do not resolve the switching current and can lead to incorrect estimation of the switching current using the peak finder method. Peak extraction is also a problem at negative gate voltages where the supercurrent magnitude is suppressed. Also, spurious dips or peaks in Rn due to charge jumps may cause an unexpected increase or decrease in $I_{sw}R_n$ values.

\section*{NS Device}
We also fabricated and measured one NS device in this study. The normal metal lead was formed by entirely covering Sn on one side of the junction with Ti/Au. Fig S5(a) shows an SEM image of the measured device. A top gate was fabricated at the boundary between the bare semiconductor and the semi-super segment to modulate the tunnel coupling between them. We performed pseudo 4-terminal $V_{bias}$ measurements for studying the superconducting gap and the magnetic field dependence. In Fig S5(b), we show the differential conductance as a function of the applied source drain bias and the top gate voltage. For this device, we obtain a superconducting gap $\Delta \approx$  $560 \mu V$, comparable to the bulk gap of Sn. This value is also in good agreement with the gap value obtained from the SNS junctions measured. In Fig S5(c), we look at the differential conductance as a function of $V_{bias}$ and the magnetic field parallel to the nanowire. We see that the gap remains open till 1.5 T. In Fig S5(d), (e) and (f), we show $V_{bias}$ vs $V_g$ plots at different magnetic fields, with the gap softening at higher magnetic fields.

\begin{figure*}
\includegraphics[width=6in]{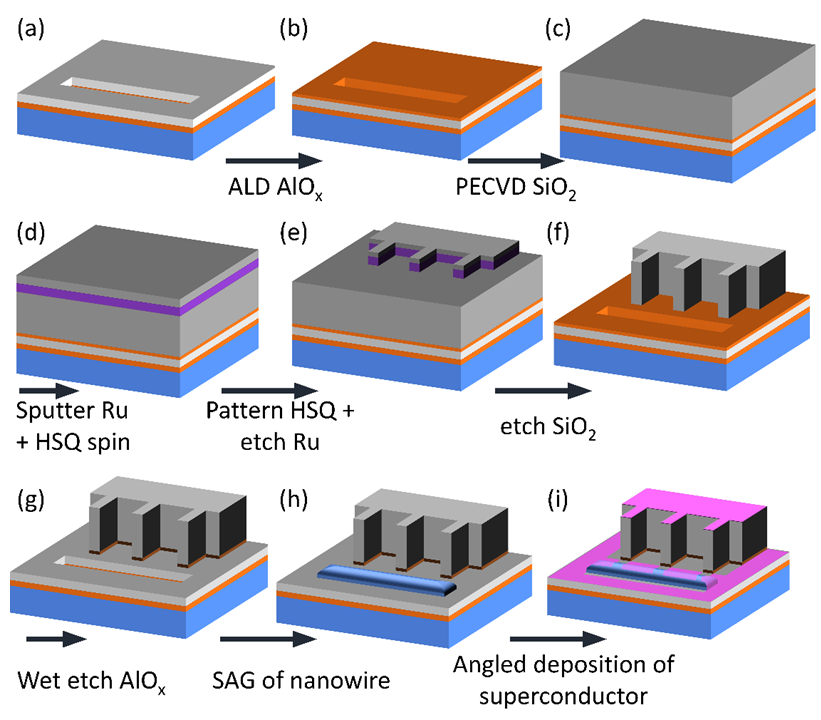} 
\centering
\caption{\label{fig:wide}(A-G)Fabrication steps to form shadow wall selective area growth templates (H) After growth of nanowire (I) After tin shadow deposition (tin is marked in pink)}
\end{figure*}

\begin{figure*}
\includegraphics[width=6in]{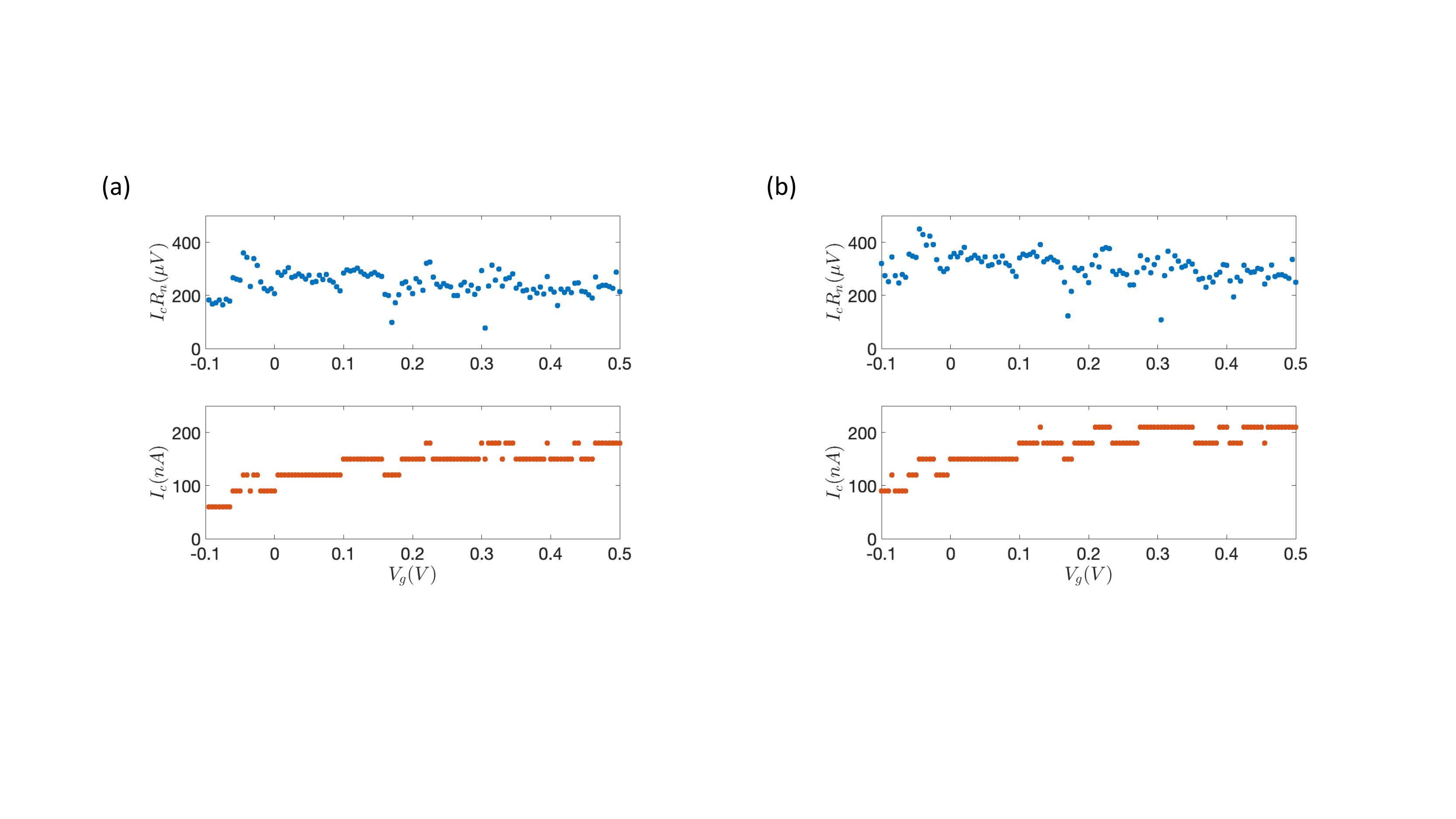} 
\centering
\caption{Comparison of $I_c$ and $I_{c}R_{n}$ product as a function of top gate voltage $V_g$ using (a) threshold method, (b) peak finder method for device BL2.5, Chip AGCBE100-2.}
\end{figure*}

\begin{figure*}
\includegraphics[width=6in]{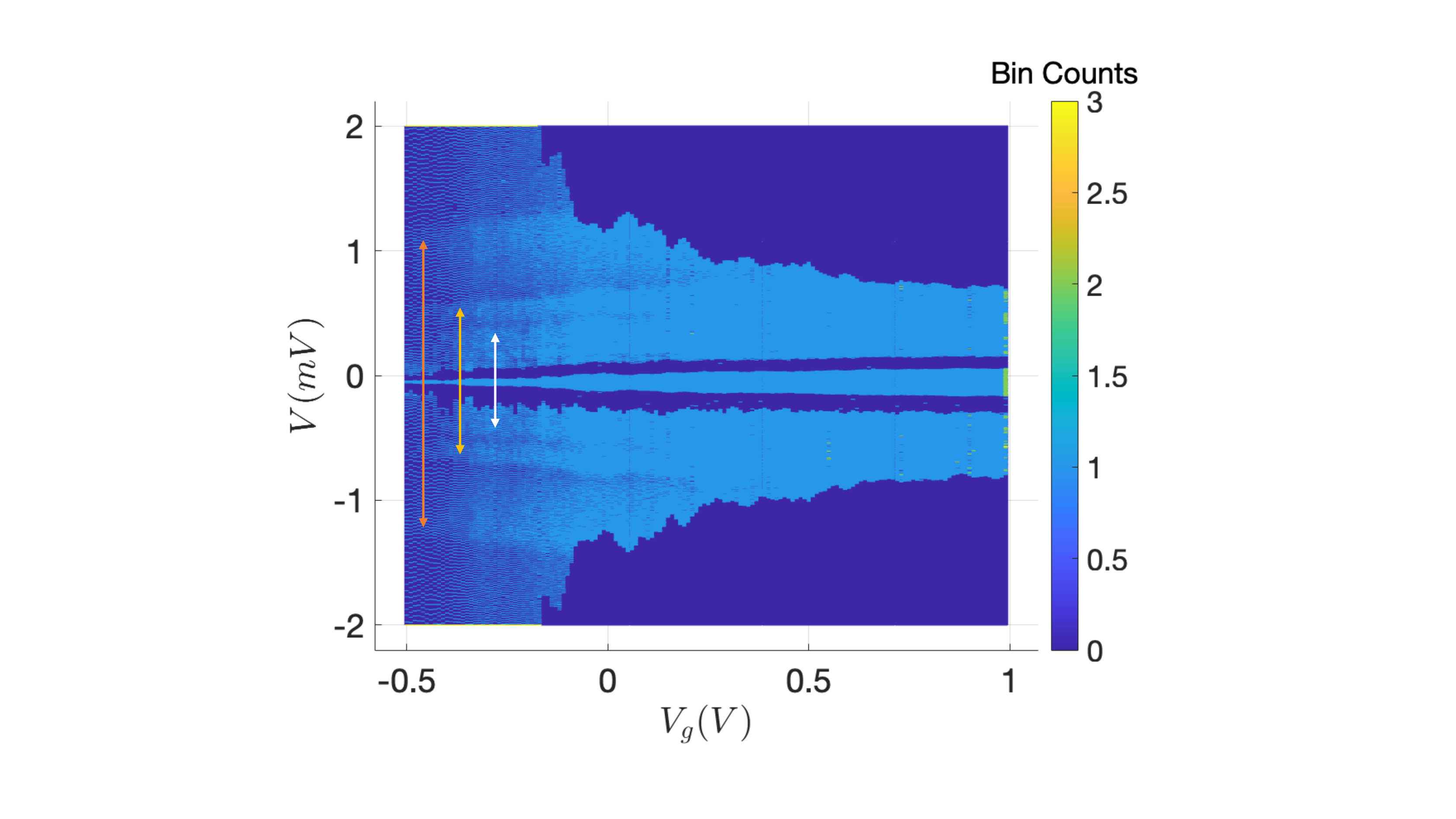} 
\centering
\caption{Fig. 2(C) plotted as a histogram by binning the measured voltage and counting the number of differential resistance points that fall within each bin (bin counts). MARs, which are constant voltage features, appear as horizontal lines with higher bin counts. The orange, yellow and white double-headed arrows denote the 4$\Delta$, 4$\Delta$/2, 4$\Delta$/3 reflections respectively, $\Delta \approx 550 \mu V$.}
\end{figure*}

\begin{table*}[t]
\begin{ruledtabular}
\begin{tabular}{p{3cm}p{2cm}p{1.5cm}p{2cm}p{2.5cm}}
Device Name& Chip Name& $\Delta$& $I_{sw}R_n$& Critical $B_{||}$ from $I_{bias}$ data\\
\hline
BL2.4 (Figs 2 and 3 in the main text)& AGCBE100-2& 560 $\mu V$& 300-600 $\mu V$& 1 T\\
\\
BL2.5& AGCBE100-2& 540 $\mu V$& 150-400 $\mu V$& 1.6 T\\
\\
BR2.4& AGCBE100-2& 585 $\mu V$& 200-500 $\mu V$& 0.9 T\\
\\
BR2.2& AGCBE100-3& 415 $\mu V$& 10-110 $\mu V$& 1.2 T\\
\\
BR2.7& AGCBE100-3& 506 $\mu V$& < 120 $\mu V$& 1.71 T\\
\\
BR2.1& AGCBE100-3& 460 $\mu V$& -& 1.5 T\\
\end{tabular}
\end{ruledtabular}
\caption{Summary of the induced gap $\Delta$, $I_{sw}R_n$ products and critical magnetic field values obtained from all the SNS junctions measured in this study.}
\end{table*}

\begin{figure*}
\includegraphics[width=6in]{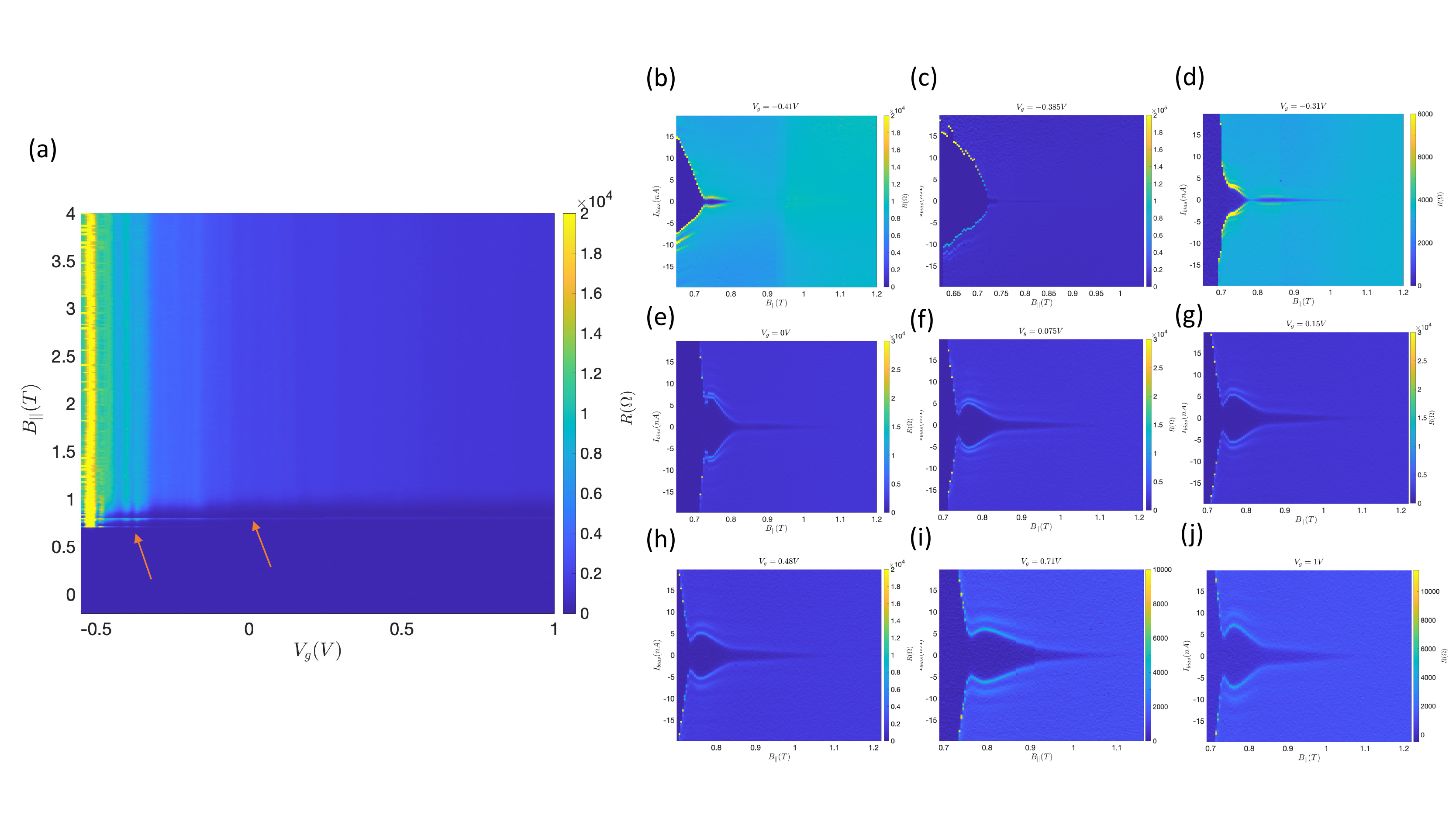} 
\centering
\caption{Additional magnetic field data for the device shown in Fig 2 : (a) $B_{||}$ vs $V_g$ using ac lockin excitation of 1 nA. Two constant magnetic field features are observed at $B_{||} = 0.73 T$ and $0.8 T$ (denoted by orange arrows). (b-j) $I_{bias}$ vs $B_{||}$ at different gate voltages. The gate voltages are noted at the top of each panel. }
\end{figure*}

\begin{figure*}
\includegraphics[width=6in]{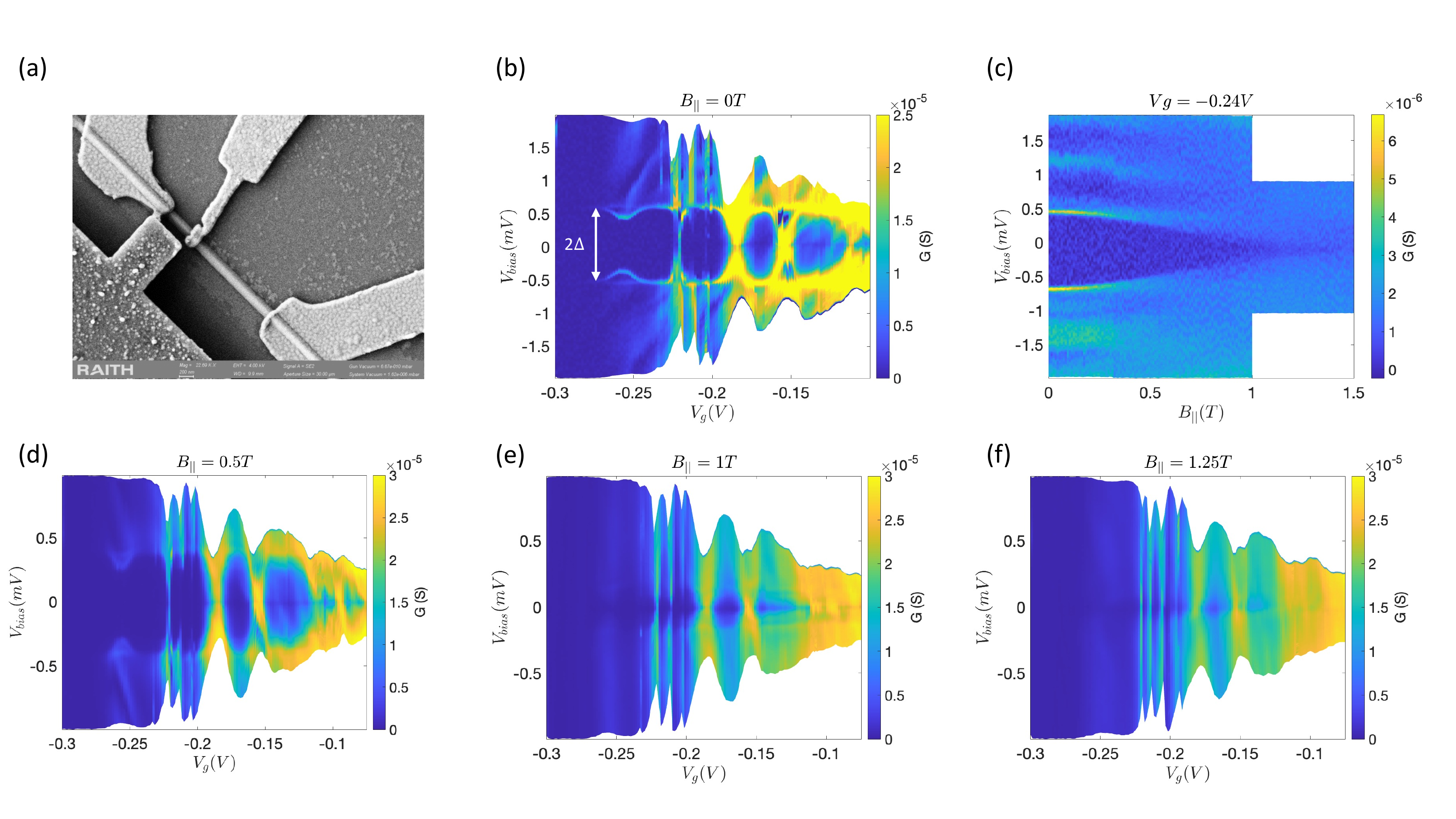} 
\centering
\caption{NS Device : (a) SEM image of the device with the source drain leads and top gate at the boundary between the bare InAs and Sn covered InAs segments, (b) Differential conductance as a function of $V_{bias}$ and $V_g$ at zero in-plane parallel magnetic field with $\Delta \approx 560 \mu V$, (c) Differential conductance as a function of $V_{bias}$ and parallel magnetic field $B_{||}$ at $V_g$ = -0.24 V, (d-f) Differential conductance as a function of $V_{bias}$ and $V_g$ at $B_{||}$ = 0.5 T, 1 T and 1.25 T respectively.}
\end{figure*}

\begin{figure*}
\includegraphics[width=6in]{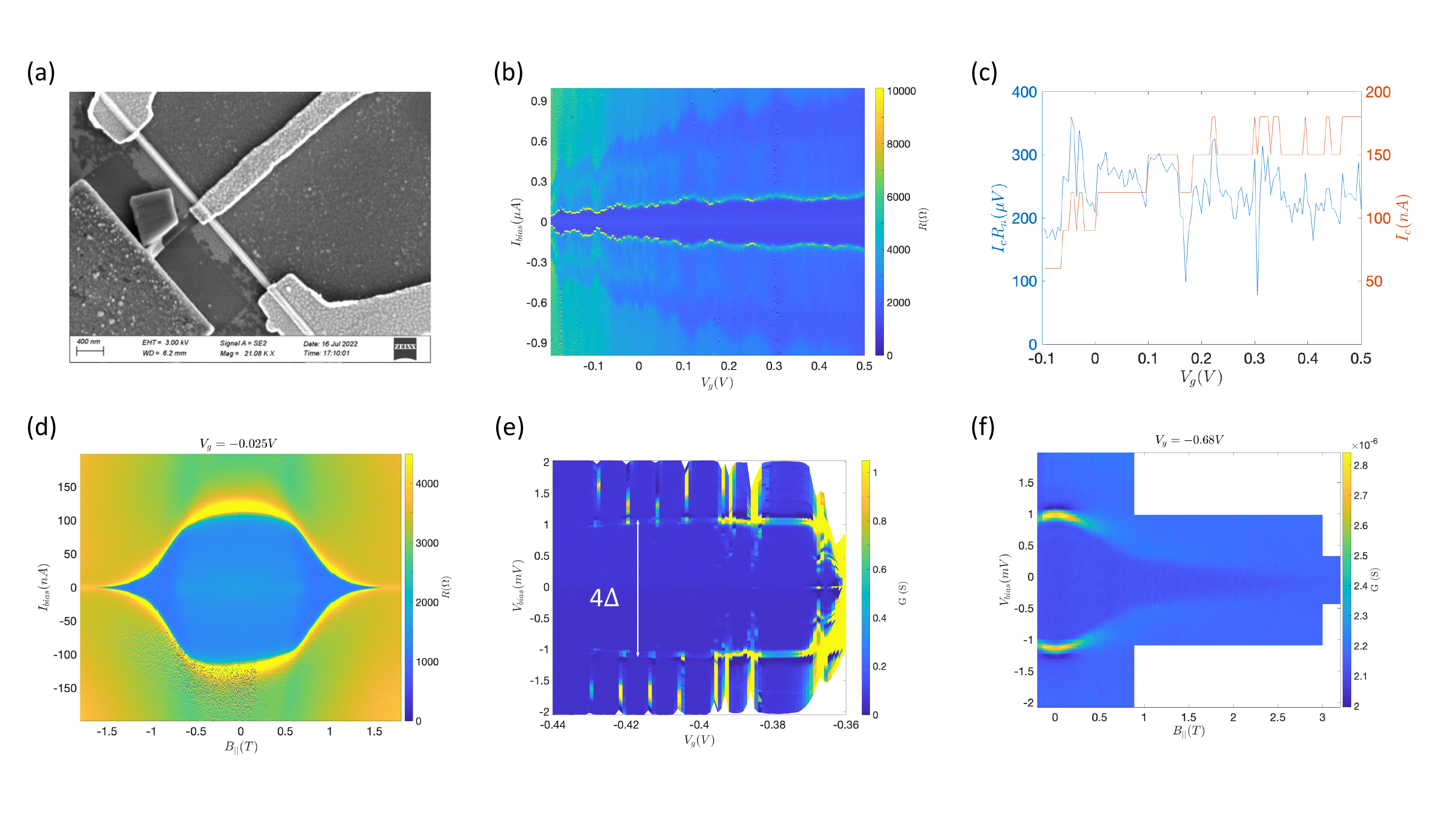} 
\centering
\caption{Data from another SNS junction measured in this study (BL2.5) : (a) SEM image of the device, (b) Differential resistance as a function of $I_{bias}$ and top gate voltage $V_g$, (c) $I_c$ and $I_{c}R_{n}$ product calculated using the threshold method as a function of $V_g$, (d) Differential resistance as a function of $I_{bias}$ and axial magnetic field $B_{||}$ at $V_g$ = -0.025 V. Critical magnetic field = 1.6 T, (e) Differential conductance as a function of $V_{bias}$ and $V_g$. The difference between the two finite bias conductance peaks is $4\Delta$, which gives an apparent superconducting gap $\Delta = 540 \mu V$, which agrees well with the bulk gap of Sn, (f) Differential conductance as a function of $V_{bias}$ and $B_{||}$ at $V_g = -0.68 V$. We see that the gap persists beyond 3 T, with the gap softening at higher magnetic fields.}
\end{figure*}

\begin{figure*}
\includegraphics[width=6in]{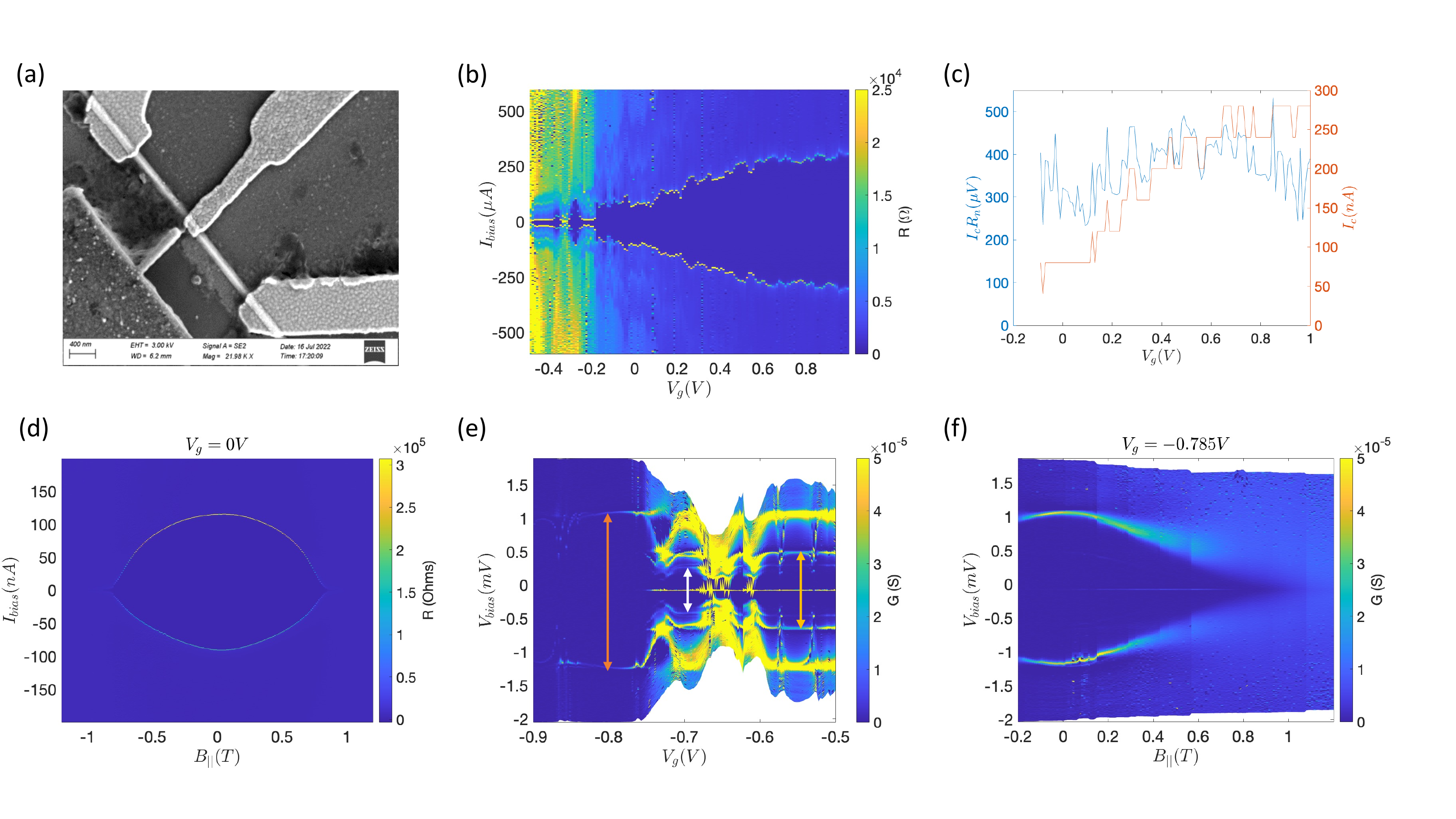} 
\centering
\caption{Data from a third SNS junction measured in this study (BR2.4) : (a) SEM image of the device, (b) Differential resistance as a function of $I_{bias}$ and top gate voltage $V_g$, (c) $I_c$ and $I_{c}R_{n}$ product calculated using the threshold method as a function of $V_g$, (d) Differential resistance as a function of $I_{bias}$ and axial magnetic field $B_{||}$ at $V_g$ = 0 V. Critical magnetic field = 0.9 T, (e) Differential conductance as a function of $V_{bias}$ and $V_g$. The orange, yellow and white double-headed arrows correspond to the first, second and third order MAR peaks respectively. The apparent superconducting gap $\Delta = 585 \mu V$, which agrees well with the bulk gap of Sn, (f) Differential conductance as a function of $V_{bias}$ and $B_{||}$ at $V_g = -0.785 V$. We see that the gap persists beyond 1 T, which is higher than the critical field obtained from (d).}
\end{figure*}

\begin{figure*}
\includegraphics[width=6in]{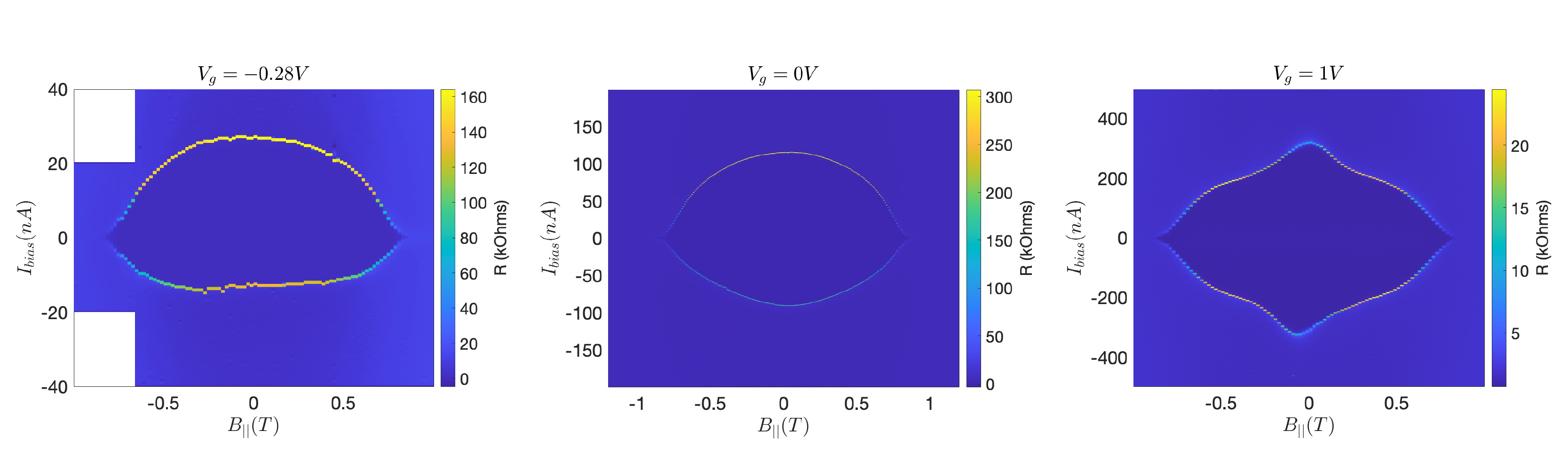} 
\centering
\caption{Additional magnetic field data for the device shown in Fig S7 (BR2.4) at (a) $V_g = -0.28 V$, (b) $V_g = 0 V$ and (c) $V_g = 1 V$}
\end{figure*}

\begin{figure*}
\includegraphics[width=6in]{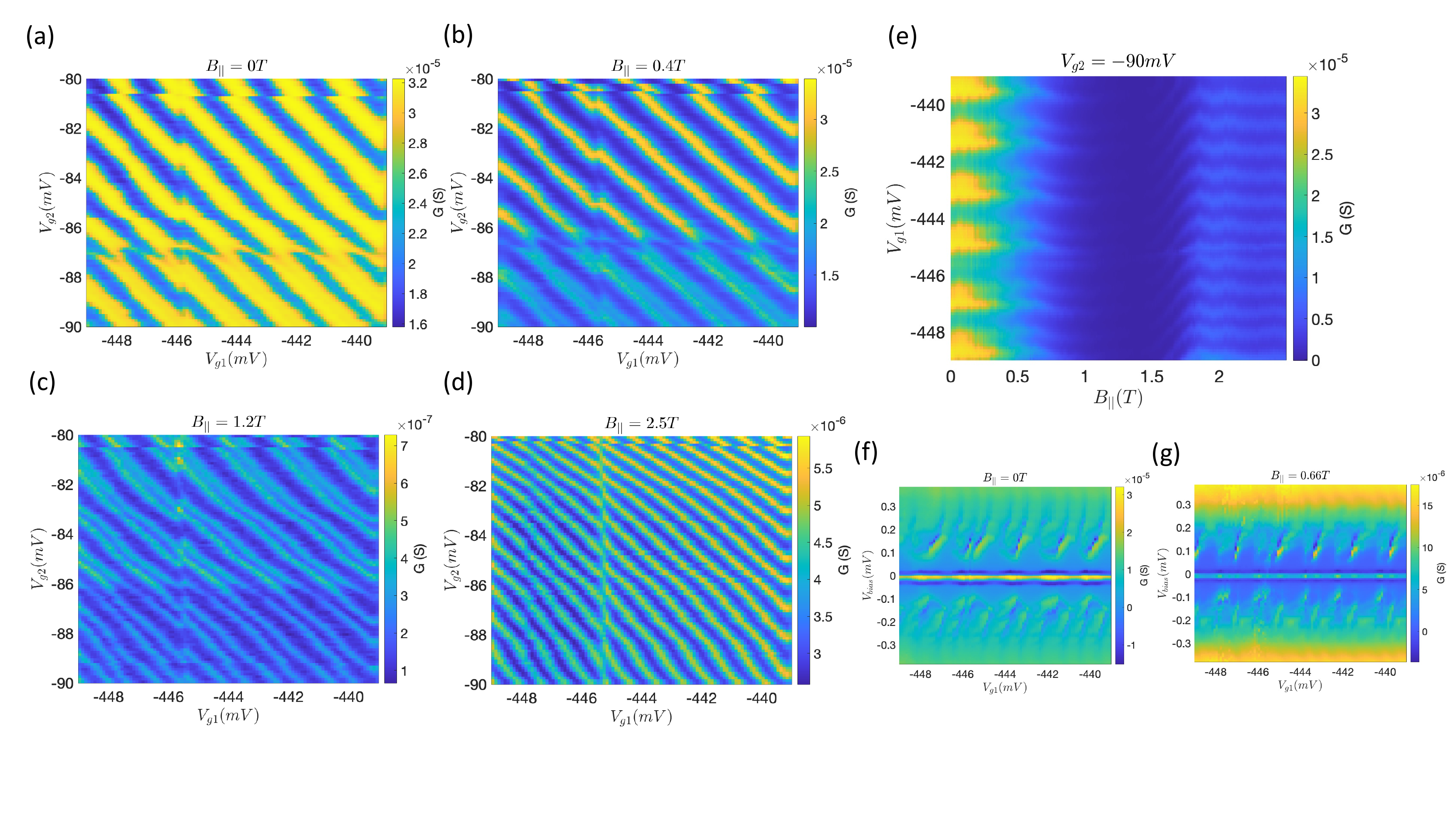} 
\centering
\caption{Additional data for the island device presented in Fig 4 : (a-d) $V_{g1}$ vs $V_{g2}$ at $B_{||}$ = 0 T, 0.4 T, 1.2 T and 2.5 T respectively. For all $V_{g1}$ in the range presented here, we see an even ground state for $V_{g2}$ > -86 mV and an even-odd ground state for $V_{g2}$ <-86 mV at zero magnetic field. Even-odd ground state at $B_{||}$ = 0 T has also been reported previously in InSb/Al and InAs/Al systems [ref]. At very high magnetic fields, superconductivity is destroyed and single electron transport sets in which causes a periodicity doubling of the Coulomb peaks. (e) Differential conductance as a function of $V_{g1}$ and $B_{||}$ at $V_{g2}$ = -90 mV i.e. staring at the even-odd ground state. (f,g) Differential conductance as  a function of $V_bias$ and $V_{g1}$ at $V_{g2}$ = -90 mV at $B_{||}$ = 0 T and 0.66 T respectively.}
\end{figure*}

\end{document}